\documentstyle[amssymb,12pt]{article}

\begin{document}

\title{On modification of the Newton's law of gravity at very large distances }
\author{A.A. Kirillov \\
{\em Institute for Applied Mathematics and Cybernetics} \\
{\em 10 Uljanova Str., Nizhny Novgorod, 603005, Russia}\\
e-mail: kirillov@unn.ac.ru \and D. Turaev \\
{\em Weierstrass Institute for Applied Analysis and Stochasticity} \\
{\em Mohrenstrasse 39, D-10117, Berlin, Germany} \\
e-mail: turaev@wias-berlin.de}
\date{}
\maketitle

\begin{abstract}
We discuss a Modified Field Theory (MOFT) in which the number of fields can
vary. It is shown that when the number of fields is conserved MOFT reduces
to the standard field theory but interaction constants undergo an additional
renormalization and acquire a dependence on spatial scales. In particular,
the renormalization of the gravitational constant leads to the deviation of
the law of gravity from the Newton's law in some range of scales $r_{\min
}<r<r_{\max }$, in which the gravitational potential shows essentially
logarithmic $\sim \ln r$ (instead of $1/r$) behavior. In this range, the
renormalized value of the gravitational constant $G$ increases and at scales 
$r>r_{\max }$ acquires a new constant value $G^{\prime }\sim Gr_{\max
}/r_{\min }$. From the dynamical standpoint this looks as if every point
source is surrounded with a halo of dark matter. It is also shown that if
the maximal scale $r_{\max }$ is absent, the homogeneity of the dark matter
in the Universe is consistent with a fractal distribution of baryons in
space, in which the luminous matter is located on thin two-dimensional
surfaces separated by empty regions of ever growing size.
\end{abstract}

  \pagebreak

\section{Introduction}

It is well\ established that dark matter gives the leading contribution to
the matter density of the Universe (e.g., see \cite{dm}). Apart from some
phenomenological properties of the dark matter (e.g., it is non-baryonic,
cold, etc.) the problem of its nature remains still open. Particle physics
suggests various hypothetical candidates for dark matter. However, while we
do not observe such particles in direct laboratory experiments there remains
the possibility to avoid or replace the dark matter paradigm. The best known
attempt of such kind is represented by the phenomenological algorithm
suggested by Milgrom \cite{mil}, the so-called MOND (Modified Newtonian
Dynamics). This algorithm suggests replacing the Newton's law of gravity in
the low acceleration limit $g\ll a_{0}$ with $g_{{\rm MOND}}\sim \sqrt{ga_{0} 
}$ (where $g$ is the gravitation acceleration and $a_{0}$ is a fundamental
acceleration $a_{0}\sim 2\times 10^{-8}cm/\,s^{2}$). This, by construction,
accounts for the two observational facts: the flat rotation curves of
galaxies and the Tully-Fisher relation $L_{gal}\propto v_{c}^{4}$ which
gives $M_{gal}$ $\propto L_{gal}\propto v_{c}^{4}$ (where $L_{gal}$, $M_{gal}
$, and $v_{c}$ are the galaxy's luminosity, mass, and rotation velocity
respectively). The MOND was shown to be successful in explaining properties
of galaxies and clusters of galaxies \cite{Sand} and different aspects of
MOND attract the more increasing attention, e.g., see Refs.\cite{Mc,L,SW}.
However, MOND presumes a nonlinear regime (e.g., at low accelerations the
force $F\propto \sqrt{M}$) and  was criticized in Ref. \cite{SW}.

A more conservative model was suggested in Ref. \cite{kin} which presumes
the existence of an additional attraction between baryons with logarithmic
potential 
\begin{equation}
U\propto \Lambda b_{1}b_{2}\ln \left( r\right)  \label{ln}
\end{equation}
where $b$ is the baryon number and $\Lambda $ defines a characteristic scale 
$r_{0}\sim 1/\Lambda \sim 5kpc$ on which this potential starts to dominate
over gravity. This model contains basic features of MOND (at least roughly)
but fails when confronting with gravitational lensing by clusters. To
explain lensing this extra force must act like gravity and, in fact, be
gravity. We note that if this additional potential cuts off at very large
distances, the effects of the extra potential will not, in fact, be
distinguishable from that of dark matter

It appears that the standard particle physics does not possess fields able
to produce interactions which has a range of scales with logarithmic
behavior. In the present paper we, however, show that the logarithmic
potentials appear naturally in the so-called Modified Field Theory (MOFT)
suggested in Ref. \cite{k99} to account for the spacetime foam effects.

It was suggested in Ref. \cite{k99} that nontrivial topology of space should
display itself in the multivalued nature of all observable fields, i.e. the
number of fields should be a dynamical variable. The argument is that in the
case of general position an arbitrary quantum state mixes different
topologies of space. From the other side, any measurement of such a state
should be carried out by a detector which obeys classical laws and,
therefore, the detector introduces a background space of a particular
topology (of course, on the classical level the topology is always defined
and does not change). This means that the topology of space must not be a
direct observable, and the only chance to keep the information on the
topology is to allow all the fields (which are specified on the background
space) to be multivalued.

The corresponding extension of the standard quantum field theory was
developed in Ref. \cite{k99} and it was proposed there that effects related
to the multivalued nature of the physical fields could reveal themselves at
large spatial scales. In the present paper we show that the logarithmic
behavior of the gravitational potential at large distances may indeed appear
as a result of nontrivial properties of the vacuum state in MOFT.\ In fact,
analogous modifications hold for all fields and, in particular, for the
Coulomb potential. In this sense at large scales we should observe not only
dark matter, but dark charges of all sorts as well.

We show that in the case when the number of fields is conserved MOFT reduces
to the standard field theory in which interaction constants (e.g., the
gravitational and the fine structure constants) undergo an additional
renormalization and, as a consequence, may acquire a dependence on spatial
scales (observational limits on the scale-dependence of the gravitational
constant have been already considered, e.g., in Ref.\cite{ob}). From the
formal standpoint such a renormalization looks as if particles lose their
point-like character and acquire an additional distribution in space, i.e.,
each particle turns out to be surrounded with a ``dark halo''.

The density distribution of the halo follows properties of the vacuum state
which formed during the quantum period in the evolution of the Universe. At
the moment, we do not have an exact model describing the formation of
properties of the vacuum and, therefore, our consideration of the vacuum
structure has a phenomenological character. Namely, we assume that upon the
quantum period of the Universe the matter was thermalized with a very high
temperature. Then, as the temperature dropped during the early stage of the
evolution of the Universe, the topological structure of the space has
tempered and the subsequent evolution resulted only in the cosmological
shift of the physical scales. We show that in MOFT such kind of assumptions
leads almost immediately to the logarithmic growth of the gravitational
potential in some range of scales.

\section{Modified Field Theory}

Let $M$ be a background basic space. Let us specify an arbitrary field $ 
\varphi $ on it. We suppose that the action for the field can be presented
in the form (for the sake of simplicity we consider the case of linear
perturbations only) 
\begin{equation}
S=\int \limits_{M}d^{4}x\left( -\frac{1}{2}\varphi \widehat{L}\varphi
+\alpha J\varphi \right) ,  \label{act1}
\end{equation}
where $\widehat{L}=\widehat{L}\left( \partial \right) $ is a differential
operator (e.g., in the case of a massive scalar field $\widehat{L}\left(
\partial \right) =\partial ^{2}+m^{2}$), $J$ is an external current, which
is produced by a set of point sources ($J=\sum J_{k}\delta \left( x-x_{k}(s)
\right) $, where $x_{k}\left( s\right) $ is a trajectory of a source), and $ 
\alpha $ is the value of the elementary charge for sources. Thus, the field $ 
\varphi $ obeys the equation of motion 
\begin{equation}
-\widehat{L}\varphi +\alpha J=0.
\end{equation}
We note that such a structure is valid for perturbations in gauge theories ($ 
\varphi =\delta A_{\mu }$, where $\alpha $ is the gauge charge) and in
gravity ($\varphi =$ $l_{pl}\delta g_{\mu \nu }$, where $\alpha =l_{pl}$ is
the Planck length).

In the Modified Field Theory we admit that the number of fields is a
variable, therefore we replace the field $\varphi $ with a set of fields $ 
\varphi ^{a}$, $a=0,1,...N\left( x\right) $. In this manner, we introduce an
additional variable $N\left( x\right) $ which in MOFT plays the role of an
operator of the number of fields. Thus, the total action assumes the
structure 
\begin{equation}
S=\int \limits_{M}d^{4}x\sum_{a=0}^{N\left( x\right) }\left( -\frac{1}{2} 
\varphi ^{a}\widehat{L}\varphi ^{a}+\alpha J\varphi ^{a}\right) ,
\label{act2}
\end{equation}
where the number of fields $N\left( x\right) $, in general, depends on the
position in the background space $M$ and, therefore, the sum stands inside
the integral over $M$. Fields $\varphi ^{a}$ are supposed to obey the
identity principle and, therefore, they equally interact with the external
current. We also note that, unlike the field $\varphi$, trajectories of
sources $x_{a}\left( s\right) $ have a single-valued character, while the
sum over sources automatically accounts for possible variations in their
number, which may appear in processes of topology changes.

It is easy to see that the main effect of the introduction of the number of
identical fields is the renormalization of the charge (the constant $\alpha $ 
). For example, let us consider the simplest case when $N\left( x\right) $
is a constant: $N\left( x\right) =N_{0}=const$. We introduce a new set of
fields as follows 
\begin{equation}
\varphi ^{a}=\frac{\widetilde{\varphi }\left( x\right) }{\sqrt{N_{0}}} 
+\delta \varphi ^{a},\,\;\;\sum_{a}\delta \varphi ^{a}=0  \label{eff}
\end{equation}
where $\widetilde{\varphi }$ is the effective field \cite{k99} 
\begin{equation}
\widetilde{\varphi }\left( x\right) =\frac{1}{\sqrt{N_{0}}} 
\sum_{a=0}^{N_{0}}\varphi ^{a}\left( x\right) .
\end{equation}
Then the action splits into two parts 
\begin{equation}
S=\int\limits_{M}d^{4}x\left( -\frac{1}{2}\sum_{a}\delta \varphi ^{a} 
\widehat{L}\delta \varphi ^{a}\right) +\int_{M}d^{4}x\left( -\frac{1}{2} 
\widetilde{\varphi }\widehat{L}\widetilde{\varphi }+\widetilde{\alpha }J 
\widetilde{\varphi }\right) .  \label{efact1}
\end{equation}
The first part represents a set of free fields $\delta \varphi ^{a}$ which
are not involved into interactions between particles and, therefore, cannot
be directly observed. The second part represents the standard action for the
effective field $\widetilde{\varphi }$ with a new value for the charge $ 
\widetilde{\alpha }=\sqrt{N_{0}}\alpha $.

In general case $N\left( x\right) $ is an operator-valued function and so
will be the charge. However, the effective field and the transformation (\ref
{eff}) can be introduced in this case as well, provided we are working with
Fourier transforms, i.e. in the momentum representation, where the states of
the fields can be classified in terms of free particles. It is, of course,
quite usual in the quantum field theory, and we show in the next section
that carrying out this approach in the framework of MOFT leads indeed to a
physically meaningful theory.

In the Fourier representation action (\ref{act2}) takes the form 
\begin{equation}  \label{act3}
S=\int dtd^{3}k\sum_{a=0}^{N(k)}\left( -\frac{1}{2}\varphi _{k}^{a\ast } 
\widehat{L}(\partial_t,-ik)\varphi_{k}^{a} +\alpha J_{k}^{\ast
}\varphi_{k}\right),
\end{equation}
where $\varphi _{k}^{a}=1/\left( 2\pi \right) ^{3/2}\int \varphi ^{a}\left(
x\right) \exp \left( -ikx\right) d^{3}x$, ($a=0,1,...N\left( k\right) $),
and $N\left( k\right) $ is the operator of the number of fields in the
momentum space (it is, of course, not the Fourier transform of $N(x)$).

When $N(k)$ conserves, making the same kind of transformation as in (\ref
{eff}) we bring the action to the form 
\begin{equation}  \label{efact}
S=\int dtd^{3}k\left( -\frac{1}{2}\sum_{a}\delta \varphi _{k}^{a\ast } 
\widehat{L}\delta \varphi _{k}^{a}-\frac{1}{2}\widetilde{\varphi }_{k}^{\ast
}\widehat{L}\widetilde{\varphi }_{k}+\widetilde{\alpha }\left( k\right)
J_{k}^{\ast }\widetilde{\varphi }_{k}\right).
\end{equation}
We see that the action for the effective field $\widetilde{\varphi}$
coincides with that in the standard theory, but the charge $\widetilde{ 
\alpha }\left(k \right) =\sqrt{N\left(k\right) }\alpha $ becomes now
scale-dependent.

We recall that $N\left( k\right) $ is an operator and we, strictly speaking,
should consider an average value for the charge 
\begin{equation}
\left\langle \widetilde{\alpha }\left( k\right) \right\rangle =\left\langle 
\sqrt{N\left( k\right) }\right\rangle \alpha .  \label{charge}
\end{equation}
The homogeneity and isotropy of the Universe allows $\left\langle N\left(
k\right) \right\rangle =N_{k}\left( t\right) $ to be an arbitrary function
of $|k|$.

In a sense, the function $N_k$ characterizes the structure of the momentum
space $M^*$ (note that the structure of the basic space $M$ itself is not
specified here). If we assume (and we do so) that processes with topology
transformations have stopped after the quantum period in the evolution of
the Universe, then the structure of the momentum space conserves indeed and
the function $\left\langle N\left( k\right)\right\rangle $ depends on time
via only the cosmological shift of scales, i.e., $\left\langle N\left(
k\right) \right\rangle =N_{k\left( t\right) }$, where $k\left( t\right) \sim
1/a\left( t\right) $ and $a\left( t\right) $ is the scale factor. In this
manner, function $N_{k}$ represents some new universal characteristic of the
physical space.

\section{Vacuum state in MOFT}

In this section we describe the structure of the vacuum state for effective
fields. For simplicity, we consider a real scalar field, while
generalization to the case of spin one and spin two particles is obvious.
Consider the expansion of the field operator $\varphi $ in plane waves, 
\begin{equation}
\varphi \left( x\right) =\sum_{k}\left( 2\omega _{k}L^{3}\right)
^{-1/2}\left( c_{k}e^{ikx}+c_{k}^{+}e{\ }^{-ikx}\right) ,  \label{field}
\end{equation}
where $\omega _{k}=\sqrt{k^{2}+m^{2}}$, and $k=2\pi n/L$, with $n=\left(
n_{x},n_{y},n_{z}\right) $; for the sake of convenience, we introduce
periodic boundary conditions with a period length $L$ (when it is necessary
sums can be replaced with integrals, as $L\rightarrow \infty $, via the
usual prescription: $\sum \rightarrow \int \left( L/2\pi \right) ^{3}d^{3}k$ 
). In the case of free particles the expression for the Hamiltonian is 
\begin{equation}
H_{0}=\sum_{k}\omega _{k}c_{k}^{+}c_{k}\,\,.  \label{H}
\end{equation}

When the number of fields is variable, the set of annihilation/creation
operators $\left\{ c_{k},c_{k}^{+}\right\} $ is replaced by the expanded set 
$\left\{ c_{a,k},c_{a,k}^{+}\right\} $, where $a\in \left[ 1,\ldots N_{k} 
\right] $, and $N_{k}$ is the number of fields for a given wave number $k$.
For a free field the energy is an additive quantity, so it can be written as 
\begin{equation}
H_{0}=\sum_{k}\sum_{a=1}^{N_{k}}\omega _{k}c_{a,k}^{+}c_{a,k}\,\;.
\end{equation}
In MOFT it is supposed that the fields are identical and obey the Fermi
statistics (for motivations and more details, see Ref. \cite{k99}). Thus,
the eigenvalues of the Hamiltonian can be written straightforwardly 
\begin{equation}
\widehat{H}_{0}=\sum_{n,k}n\omega _{k}N_{n,k}\,,  \label{zu}
\end{equation}
where $N_{n,k}$ is the number of field modes with the given wave number $k$
and number of scalar particles $n$; since we assume Fermi statistics, we
should set $N_{n,k}=0$ or $1$. Assuming that upon the quantum period of the
evolution of the Universe topology transformations are suppressed, we should
require that the number of fields conserves $\sum_{n}N_{n,k}=N_{k}=const$ in
every mode. Thus, we find that the field ground state $\Phi _{0}$ is
characterized by occupation numbers 
\begin{equation}
N_{n,k}=\theta \left( \mu _{k}-n\omega _{k}\right) ,  \label{GST}
\end{equation}
where $\theta \left( x\right) $ is the Heaviside step function and $\mu _{k}$
is the chemical potential. For the spectral number of fields we get 
\begin{equation}
N_{k}=\sum_{n=0}^{\infty }\theta \left( \mu _{k}-n\omega _{k}\right) =1+ 
\left[ \frac{\mu _{k}}{\omega _{k}}\right] \,\,,  \label{N}
\end{equation}
where $[x]$ denotes the integral part of $x$. We interpret the function $ 
N_{k}$ as a geometric characteristic of the momentum space, so it should be
common for all types of Bose fields. Then assigning a specific value for the
function $N_{k}$, the expression (\ref{N}) defines the value of the
respective chemical potential $\mu _{k}$ for a given field.

The creation and annihilation operators for the effective field are
introduced as follows \cite{k99} 
\begin{equation}
\widetilde{c}_{k}=\frac{1}{\sqrt{N_{k}}}\sum_{a=1}^{N_{k}}c_{a,k},\;\;\; 
\widetilde{c}_{k}^{+}=\frac{1}{\sqrt{N_{k}}}\sum_{a=1}^{N_{k}}c_{a,k}^{+},
\label{norm}
\end{equation}
while the interaction term in (\ref{efact}) takes the form 
\begin{equation}
S_{int}=\int dt\sum_{k}\widetilde{\alpha }\left( k\right) \left( \widetilde{c 
}_{k}J_{k}+\widetilde{c}_{k}^{+}J_{k}^{+}\right)
\end{equation}
with $\widetilde{\alpha }\left( k\right) =\sqrt{N_{k}}\alpha $.

\section{Origin of the spectral number of fields}

As it was already mentioned above, the function $N_{k}$ forms during the
quantum stage of the evolution of the Universe, when processes involving
topology changes took place. It is well known that near the singularity the
evolution of the Universe is governed by a scalar field (responsible for a
subsequent inflationary phase). For the sake of simplicity we neglect the
presence of all other sorts of particles during the quantum stage and
suppose that the Universe was filled with massive scalar particles only.

Upon the quantum period the Universe is supposed to be described by the
homogeneous metric of the form 
\begin{equation}
ds^{2}=dt^{2}-a^{2}\left( t\right) dl^{2},
\end{equation}
where $a\left( t\right) $ is the scale factor, and $dl^{2}$ is the spatial
interval. It is natural to expect (at least it is the simplest possibility)
that the state of the scalar field was thermalized with a very high
temperature $T>T_{Pl}$ where $T_{Pl}$ is the Planck temperature. The state
of the field was characterized by the thermal density matrix with $\mu =0$
(for the number of fields varies) and with mean values for occupation
numbers $\left\langle N_{k,n}\right\rangle =\left( \exp \left( \frac{n\omega
_{k}}{T}\right) +1\right) ^{-1}$. On the early stage $m\ll T$, and the
temperature and the energy of particles depend on time as $T=\widetilde{T} 
/a\left( t\right)$, $k=\widetilde{k}/a\left( t\right)$. When the temperature
drops below a critical value $T_{\ast }$ , which corresponds to the moment $ 
t_{\ast }\sim t_{pl}$, topological structure (and the number of fields)
tempers. This generates the value of the chemical potential for scalar
particles $\mu \sim T_{\ast }$.

Let us neglect the temperature corrections, which are essential only at $ 
t\sim t_{\ast }$ and whose role is in smoothing the real distribution $N_{k}$ 
. Then at the moment $t\sim t_{\ast }$ the ground state of the field will be
described by (\ref{GST}) with $\mu _{k}=\mu =const\sim T_{\ast }$. During
the subsequent evolution the\ physical scales are subjected to the
cosmological shift, however the form of this distribution in the commoving
frame must remain the same. Thus, on the later stages $t\geq t_{\ast }$, we
find 
\begin{equation}
N_{k}=1+\left[ \frac{\widetilde{k}_{1}}{\Omega _{k}\left( t\right) }\right]
,\,\,  \label{NN}
\end{equation}
where $\Omega _{k}\left( t\right) =\sqrt{a^{2}\left( t\right) k^{2}+ 
\widetilde{k}_{2}^{2}}$, $\widetilde{k}_{1}\sim a_{0}\mu $, and $\widetilde{k 
}_{2}\sim a_{0}m$ ($a_{0}=a\left( t_{\ast }\right) $).

Consider now properties of the function $N_{k}$. There is a finite interval
of wave numbers $k\in \lbrack k_{\min }\left( t\right) ,k_{\max }\left(
t\right) ]$ on which the number of fields $N_{k}$ changes its value from $ 
N_{k}=1$ (at the point $k_{\max }$) to the maximal value $N_{\max }=1+\left[ 
\widetilde{k}_{1}/\widetilde{k}_{2}\right] $ (at the point $k_{\min }$).
This causes the variation of the charge values from $\alpha _{\min }=\alpha $ 
, to $\alpha _{\max }=\sqrt{N_{\max }}\alpha $. The boundary points of the
interval of $k$ depend on time and are expressed via the free
phenomenological parameters $\widetilde{k}_{1}$ and $\widetilde{k}_{2}$ as
follows 
\[
k_{\max }=\frac{1}{a\left( t\right) }\sqrt{\widetilde{k}_{1}^{2}-\widetilde{k 
}_{2}^{2}},\;\;k_{\min }=\frac{1}{a\left( t\right) }\sqrt{\widetilde{k} 
_{1}^{2}/\left( N_{\max }-1\right) ^{2}-\widetilde{k}_{2}^{2}}. 
\]
And in the wave number range $k\leq k_{\min }\left( t\right) $ the number of
fields remains constant $N_{k}=N_{\max }$.

During the later stages of the evolution of the Universe ($t\gg t_{\ast }$)
the contribution from all other fields becomes essential. However processes
involving topology transformations are suppressed and the structure of the
momentum space is described by the distribution (\ref{NN}). We note that the
real distribution can be different from (\ref{NN}), which depends on the
specific picture of topology transformations in the early Universe and
requires the construction of the exact theory (in particular, thermal
corrections smoothen the step-like distribution (\ref{NN})). However we
believe that the general features of $N_{k}$ will remain the same.

\section{The Law of gravity}

The dependence of charge values upon wave numbers leads to the fact that the
standard expressions for the Newton's and Coulomb's energy of interaction
between point particles break down. In this section we consider corrections
to the Newton's law of gravity (corrections to the Coulomb's law are
identical). The interaction constant $\alpha \sim m\sqrt{G}$ (where $m$ is
the mass of a particle), and MOFT gives $G\rightarrow G\left( k\right)
=N_{k}G$. To make estimates, we note that at the moment $t\sim t_{\ast }$
the mass of scalar particles should be small as compared with the chemical
potential (which has the order of the Planck energy), which gives $ 
\widetilde{k}_{1}\gg \widetilde{k}_{2}$. Then in the range $k_{\max }\left(
t\right) \geq k$ $\gg k_{\min }\left( t\right) $ the function $N_{k}$ can be
approximated by 
\[
N_{k}\sim 1+\left[ k_{\max }\left( t\right) /k\right]. 
\]

Consider two rest point particles with masses $m_{1}$ and $m_{2}$. Then the
Fourier transform for the energy of the gravitational interaction between
particles is given by the expression 
\begin{equation}
V\left( {\bf k}\right) =-\frac{4\pi Gm_{1}m_{2}}{\left| {\bf k}\right| ^{2}} 
N_{k},  \label{80}
\end{equation}
which in the coordinate representation is given by the integral 
\begin{equation}
V\left( r\right) =\frac{1}{2\pi ^{2}}\int\limits_{0}^{\infty }\left( V\left(
\omega \right) \omega ^{3}\right) \frac{\sin \left( \omega r\right) }{\omega
r}\frac{d\omega }{\omega }.  \label{90}
\end{equation}
From (\ref{N}) and (\ref{NN}) we find that this integral can be presented in
the form 
\begin{equation}
V\left( r\right) =-\frac{2Gm_{1}m_{2}}{\pi }\sum\limits_{n=0}^{N_{\max
}-1}\int\limits_{0}^{k_{n}}\frac{\sin \left( \omega r\right) }{\omega r} 
d\omega =-\frac{Gm_{1}m_{2}}{r}\left( 1+\sum\limits_{n=1}^{N_{\max }-1}\frac{ 
2 {Si}\left( k_{n}r\right) }{\pi }\right)  \label{NW}
\end{equation}
where $k_{n}=\frac{1}{a\left( t\right) n}\sqrt{\widetilde{k}_{1}^{2}-n^{2} 
\widetilde{k}_{2}^{2}}.$ The first term $n=0$ of the sum in (\ref{NW}) gives
the standard expression for the Newton's law of gravity, while the terms
with $n>1$ describe corrections. From (\ref{NW}) we find that in the range $ 
k_{1}r=k_{\max }r\ll 1$, $Si\left( k_{n}r\right) \sim k_{n}r$ and
corrections to the Newton's potential give the constant 
\begin{equation}
\delta V\sim -\frac{2Gm_{1}m_{2}}{\pi }\sum\limits_{n=1}^{N_{\max }-1}k_{n}.
\end{equation}
Thus, in this range we have the standard Newton's force. In the range $ 
k_{\min }r\gg 1$, we get $\frac{2}{\pi }Si\left( k_{n}r\right) \sim 1$ and
for the energy (\ref{NW}) we find 
\[
V\left( r\right) \sim -\frac{G^{\prime }m_{1}m_{2}}{r}, 
\]
where $G^{\prime }=GN_{\max }$. Thus, on scales $r\gg 1/k_{\min }$ the
Newton's law is restored, however the gravitational constant increases in $ 
N_{\max }$ times. In the intermediate range $1/k_{\min }\gg r\gg 1/k_{\max }$
the corrections can be approximated as 
\begin{equation}
\delta V\left( r\right) \sim \frac{2Gm_{1}m_{2}}{\pi }\frac{\widetilde{k}_{1} 
}{a\left( t\right) }\ln \left( \frac{\widetilde{k}_{2}}{a\left( t\right) } 
r\right) ,  \label{110}
\end{equation}
i.e., they have a logarithmic behavior.

We note, that from the dynamical point of view the modification of the
Newton's law of gravity can be interpreted as if point sources lose their
point-like character and acquire an additional distribution in space.
Indeed, let $m_{1}$ be a test particle which moves in the gravitational
field created by a point source $m_{2}$. Then, assuming the Newton's law is
unchanged, from (\ref{NW}) we conclude that the source $m_{2}$ is
distributed in space with the density 
\begin{equation}
\rho \left( r\right) =\frac{m_{2}}{2\pi ^{2}}\int\limits_{0}^{\infty }\left(
N_{k}k^{3}\right) \frac{\sin \left( kr\right) }{kr}\frac{dk}{k}=m_{2}\left(
\delta \left( \vec{r}\right) +\frac{1}{2\pi ^{2}}\sum\limits_{n=1}^{N_{\max
}-1}\frac{\sin \left( k_{n}r\right) -k_{n}r\cos \left( k_{n}r\right) }{r^{3}} 
\right) .  \label{dark0}
\end{equation}
The total mass contained within a radius $r$ is 
\begin{equation}
M\left( r\right) =4\pi \int_{0}^{r}s^{2}\rho \left( s\right) ds=m_{2}\left(
1+\frac{2}{\pi }\sum\limits_{n=1}^{N_{\max }-1}\left( Si\left( k_{n}r\right)
-\sin \left( k_{n}r\right) \right) \right) .  \label{m}
\end{equation}
Thus, in the range $r\ll 1/k_{\max }$ we find $M\left( r\right) \sim m_{2}$,
i.e., one may conclude that the gravitational field is created by a point
source with the mass $m_{2}$. However in the range $1/k_{\min }>r>1/k_{\max
} $ the mass increases $M\left( r\right) \sim m_{2}k_{\max }r$, and for $ 
r\gg 1/$ $k_{\min }$ the mass reaches the value $M\left( r\right) \sim
m_{2}k_{\max }/k_{\min }$.

We see that in MOFT the distributions of the dark matter and the actual
matter are strongly correlated (by the rule (\ref{dark0})), and the
resulting behavior of the dynamically determined mass $M(r)$ seems to agree
with the observations. We stress that the theoretical scheme of MOFT was not
invented to fit the dark matter distribution. On the contrary, the
logarithmic behavior of the effective field potentials simply appears in the
thermodynamically equilibrium state at the low temperature, as a by-product
of a non-trivial structure of MOFT vacuum.

\section{Conclusions}

In this manner we have shown that in the case when the number of fields is
conserved MOFT reduces to the standard field theory in which interaction
constants undergo a renormalization and, in general, acquire a dependence on
spatial scales. From the dynamical standpoint such a renormalization looks
as if particles lose their point-like character and acquire an additional
distribution in space, i.e., each point source is surrounded with a halo of
dark matter. This halo carries charges of all sorts and its distribution
around a point source follows properties of the vacuum. The latter forms
during the quantum period in the evolution of the Universe, and the rigorous
consideration of vacuum properties requires constructing an exact theory.

In applying to cosmological problems it is convenient to suppose that the
Newton's law of gravity remains intact, while the variation of the
gravitational constant is phenomenologically described by the presence of a
dark matter. In the simplest case properties of the vacuum and that of dark
matter can be described by two phenomenological parameters which represent
the two characteristic scales. They are the minimal scale $r_{\min }=2\pi
/k_{\max }$ on which the dark matter starts to show up (and on which the law
of gravity (\ref{NW}) starts to deviate from the Newton's law) and the
maximal scale $r_{\max }=2\pi /k_{\min }$ which defines the fraction of the
dark matter or the total increase of interaction constants $G_{\max }\approx
Gr_{\max }/r_{\min }$ (and after which the Newton's law restores). The
minimal scale $r_{\min }$ can be easily estimated (e.g., see Ref.\cite{kin})
and constitutes a few $kpc$. To get analogous estimate for the maximal scale 
$r_{\max }$ is not so easy. This requires the exact knowledge of the total
matter density $\Omega _{tot}$ for the homogeneous background and the
knowledge of the baryon fraction $\Omega _{b}$ which gives $r_{\max
}/r_{\min }\sim \Omega _{tot}/\Omega _{b}$ (where $\Omega =\rho /\rho _{cr}$
and $\rho _{cr}$ is the critical density).

If we accept the value $\Omega _{tot}\sim 1$ (which is predicted by
inflationary scenarios) and take the upper value for baryons $\Omega
_{b}\lesssim 0.03$ (which comes from the primordial nuclearsynthesis), we
find $r_{\max }/r_{\min }\gtrsim 30$. Another estimate can be found from
restrictions on parameters of inflationary scenarios. Indeed, in
inflationary models correct values for density perturbations give the upper
boundary for the mass of the scalar field $m\lesssim 10^{-5}m_{Pl}$ \ which
gives $r_{\max }/r_{\min }\gtrsim 10^{5}T_{\ast }/m_{pl}$, where $T_{\ast }$
is the critical temperature at which topology has been tempered.

From our point of view, the most interesting picture of the Universe appears
in the case when the maximal scale is absent ($r_{\max }\rightarrow \infty $ 
, or at least $r_{\max }\gg R_{H}$, where $R_{H}$ is the Hubble radius). In
this case a uniform distribution of baryons in space is consistent with
closed cosmological models only. Indeed, the number of baryons contained
within a radius $r$ in the case of a uniform distribution with a density $ 
n_{b}$ is given by $N_{b}\left( r\right) \sim n_{b}r^{3}$ and the mass of
every baryon increases according to (\ref{m}) as $m_{b}\left( r\right) \sim
m_{p}r/r_{\min }$ ($m_{p}$ is the proton mass). Thus, for the total mass
(baryons plus dark matter) contained within the radius $r$ we get $ 
M_{tot}\left( r\right) =m_{b}\left( r\right) N_{b}\left( r\right) +\delta
M\left( r\right) \geq \rho _{b}r^{4}/r_{\min }$ (where $\rho _{b}=m_{p}n_{b}$
and $\delta M\left( r\right) $ accounts for the contribution of baryons \
from the outer region which does exist according to (\ref{dark0})). This
means that the lower limit for the total density increases with the radius $ 
\rho _{tot}\sim M_{tot}/r^{3}\geq $ $\rho _{b}r/r_{\min }$ and for
sufficiently large $r$ $\sim r_{cr}$ it will reach the value $\rho
_{b}r_{cr}/r_{\min }=$ $\rho _{cr}$ , i.e., $\Omega _{tot}>1$ and such a
Universe must correspond to a closed cosmological model. Then $r_{\max }$
coincides with the radius of the Universe $r_{\max }=R$ and for the mean
value of the matter density we will find $\Omega _{tot}=\Omega _{b}R/r_{\min
}$.

We note that this does not mean that open cosmological models are forbidden
at all. This, however, means that in open models luminous matter has a
specific nonuniform distribution in space $\rho \left( x\right) $ (the total
density nevertheless is uniformly distributed, i.e., the dark matter
compensates exactly the inhomogeneity of the luminous matter). Indeed, let $ 
\Omega _{tot}$ be a constant in space and let it be of order $1$. Then the
total mass $M_{tot}\left( r\right) $ $=$ $m_{b}\left( r\right) N_{b}\left(
r\right) $ $+$ $\delta M\left( r\right) $ contained within a radius $r$
behaves as $\sim r^{3}\rho _{tot}$ and, therefore, the number of baryons
should follow the law $N_{b}\left( r\right) $ $\leq $ $r^{2}r_{\min }\rho
_{tot}/m_{p}$. Such a law for the number of baryons can be achieved when the
luminous matter is located on thin two-dimensional surfaces separated by
empty regions of ever growing size (i.e., baryons have a kind of a fractal
distribution in space). This, in fact, is consistent with the observed
picture of the Universe at large scales $r\gtrsim 100Mpc$.

In this manner in open models the mean density of baryons depends on the
scale of averaging out $\rho _{b}\left( r\right) =\rho _{tot}r_{\min }/r$
and at the Hubble scale $r\sim R_{H}$ we find $\Omega _{b}/\Omega _{tot}\sim
r_{\min }/R_{H}$. If we consider larger scales we find $\rho _{b}\rightarrow
0$ as $r\rightarrow \infty $, i.e., the expansion of the Universe is
governed by dark matter alone.

In conclusion we point out that all the interaction constants (the
gravitational constant $G$ the electron charge $e$, gauge charges) depend on
time via the cosmological shift of scales $r_{\min }\left( t\right) $ and $ 
r_{\max }\left( t\right) $ which may give rise to a number of interesting
processes in the early Universe. We do not discuss here all the
possibilities but just point out to them which, in general, may be used to
rule out or, better, to confirm the theory suggested.

\end{document}